\documentclass[12pt]{article}
\usepackage{epsfig}
\def\be{\begin{equation}}
\def\ee{\end{equation}}
\def\bea{\begin{eqnarray}}
\def\eea{\end{eqnarray}}
\usepackage{graphicx}

\catcode`\@=11
\def\lsim{\mathrel{\mathpalette\@versim<}}
\def\gsim{\mathrel{\mathpalette\@versim>}}
\def\@versim#1#2{\vcenter{\offinterlineskip
\ialign{$\m@th#1\hfil##\hfil$\crcr#2\crcr\sim\crcr } }}
\catcode`\@=12
\usepackage{axodraw}

\catcode`@=12
\begin{document}
\thispagestyle{empty}
\begin{flushright}
UCRHEP-T467\\
April 2009\
\end{flushright}
\vspace{0.8in}
\begin{center}
{\LARGE \bf Radiative Inverse Seesaw: Verifiable\\ 
New Mechanism of Neutrino Mass\\}
\vspace{1.2in}
{\bf Ernest Ma\\}
\vspace{0.2in}
{\sl Department of Physics and Astronomy, University of California,\\
Riverside, California 92521, USA\\}
\end{center}
\vspace{1.2in}
\begin{abstract}\
If the canonical seesaw mechanism alone is responsible for neutrino mass, 
i.e. $m_\nu \simeq -m_D^2/m_N$, it can neither be proved nor disproved at 
the TeV energy scale.  A new verifiable mechanism of neutrino mass is 
proposed, using the $inverse$ seesaw, with new physics at the TeV scale, 
such that $m_\nu \simeq m_D^2 \epsilon_L/m_N^2$, where $\epsilon_L$ is a 
two-loop effect.  Dark-matter candidates also appear naturally.
\end{abstract}

\newpage
\baselineskip 24pt

\noindent \underline{\it Introduction}~:~ Neutrinos have mass, but the 
mechanism for it to occur remains a topic of theoretical study.  The reason 
is that, unlike other charged fermions such as the electron or the quarks, 
the neutrinos are electrically neutral, and could have either Dirac or 
Majorana masses or both.  The prevalent thinking is that in addition to the 
left-handed neutrino $\nu_L$ in the electroweak lepton doublet $(\nu,l)_L$ 
of the standard $SU(3)_C \times SU(2)_L \times U(1)_Y$ gauge model of particle 
interactions, there is also a right-handed neutrino $N_R$ (for each of the 
three families of quarks and leptons), which is a singlet.  Thus it has no 
gauge interactions, and couples only to the Higgs doublet $\Phi = (\phi^+,
\phi^0)$, i.e. $f \bar{N}_R (\nu_L \phi^0 - l_L \phi^+)$, so that a Dirac mass 
is obtained as $\phi^0$ acquires a vacuum expectation value $v = \langle 
\phi^0 \rangle$, linking $\nu_L$ with $N_R$.  If this is the only allowed 
additional term, then $N_R$ is simply $\nu_R$, i.e. $\nu$ is a four-component 
Dirac spinor with mass $m_D = fv$, and additive lepton number $L$ is conserved.
However, since $N_R$ is a singlet, it should be allowed a Majorana mass $m_N$. 
Hence the $2 \times 2$ mass matrix spanning $\bar{\nu}_L$ and $N_R$ is given by
\begin{equation}
{\cal M}_{\nu N} = \pmatrix{0 & m_D \cr m_D & m_N},
\end{equation}
with eigenvalues $m_{1,2} = m_N/2 \mp \sqrt{(m_N/2)^2 + m_D^2}$.  It is 
customarily assumed that $m_D << m_N$, in which case $m_1 \simeq -m_D^2/m_N$ 
and $m_2 \simeq m_N$.  This is the famous canonical seesaw mechanism 
\cite{seesaw} and explains why $m_1$ (which is then renamed $m_\nu$) is so 
small.  However, the mixing between $\nu_L$ and $N_R$ is $|m_D/m_N| \simeq 
\sqrt{|m_\nu/m_N|}$ which is at most $10^{-6}$ (for $m_\nu = 1$ eV and $m_N = 1$ 
TeV) and precludes any observable effect in support of this hypothesis. 
If this is the correct mechanism of neutrino mass, it may never be proved 
or disproved \cite{m05}.

\noindent \underline{\it Inverse seesaw}~:~ There are other mechanisms of 
neutrino mass \cite{m98} and some may be verifiable at the TeV scale 
\cite{mrs00,m01}.  In this paper, a new mechanism is proposed, where the 
origin of neutrino mass is radiative and suppressed by the inverse seesaw 
\cite{ww83,mv86,m87,dv05,kk07,m09-1} due to new physics at the TeV scale.  
The basic framework of the inverse seesaw is to extend Eq.~(1) to include 
one additional singlet $N_L$, so that the resulting $3 \times 3$ mass matrix 
spanning $\bar{\nu}_L$, $N_R$, and $\bar{N}_L$ becomes \cite{m09-1}
\begin{equation}
{\cal M}_{\nu N} = \pmatrix{0 & m_D & 0 \cr m_D & m_R & m_N \cr 0 & m_N & m_L}.
\end{equation}
Thus $m_D$ is the usual Dirac mass linking $\nu_L$ with $N_R$ through $\langle 
\phi^0 \rangle$, and $m_N$ is an invariant Dirac mass, whereas $m_R$ and $m_L$ 
are Majorana mass terms.  If $m_{R,L} = 0$, then the linear combination 
$(m_D \nu_L + m_N N_L)/\sqrt{m_D^2 + m_N^2}$ will combine with $N_R$ to form 
a Dirac fermion of mass $\sqrt{m_D^2 + m_N^2}$ and the orthogonal combination 
\begin{equation}
\nu_1 = {m_N \nu_L - m_D N_L \over \sqrt{m_D^2 + m_N^2}}
\end{equation}
remains massless.  Additive lepton number $L$ is conserved in this case.  
This limit allows one to argue that $m_{R,L}$ should be small, because in 
their absence, the symmetry of the resulting theory is enlarged, i.e. from 
$(-)^L$ to $L$.  In all previous applications, these small parameters are 
simply put in by hand.  Here it will be shown how they may only be radiatively 
generated and must therefore be small.

Renaming $m_{R,L}$ as $\epsilon_{R,L}$, and using $\epsilon_{R,L} << m_D, m_N$, 
the eigenvalues of Eq.~(2) are:
\begin{eqnarray}
m_1 &=& {m_D^2 \epsilon_L \over m_N^2 + m_D^2}, \\ 
m_2 &=& \sqrt{m_N^2 + m_D^2} + {\epsilon_R \over 2} + {m_N^2 \epsilon_L \over 
2(m_N^2+ m_D^2)}, \\ 
m_3 &=& -\sqrt{m_N^2 + m_D^2} + {\epsilon_R \over 2} + {m_N^2 \epsilon_L \over 
2(m_N^2+ m_D^2)},
\end{eqnarray}
where $m_1$ is now an inverse seesaw neutrino mass.  It is small because 
$\epsilon_L$ is small, without requiring $m_N$ to be excessively large. 
For example, let $m_D \sim 10$ GeV, $m_N \sim 1$ TeV, and $\epsilon_L \sim 10$ 
keV, then $m_1 \sim 1$ eV.  Note that $\nu_1$ is again given by Eq.~(3) to a 
very good approximation.  The mixing of $\nu_L$ with $N_R$ remains very small, 
i.e. $m_D \epsilon_L/(m_N^2 + m_D^2)$, but the mixing of $\nu_L$ with $N_L$ 
is $m_D/m_N$, which may be large enough to be observed, as unitarity violation 
in future neutrino experiments \cite{bgw04,xz08,abf09,moz09,abfl09,r09}, as 
well as lepton flavor violation.

\noindent \underline{\it $U(1)_\chi$ extension of the SM}~:~ To enforce the 
form of Eq.~(2) where $m_{R,L}$ are necessarily radiative, a gauge extension 
of the SM is recommended.  As a concrete example, consider the breaking of
\begin{equation} 
SO(10) \to SU(5) \times U(1)_\chi \to SU(3)_C \times SU(2)_L \times U(1)_Y 
\times U(1)_\chi.
\end{equation}
This is simply achieved with a Higgs scalar multiplet \underline{45} which 
decomposes as 
\begin{equation}
\underline{45} = (\underline{1},0) + (\underline{10},1) + (\underline{10}^*,-1) 
+ (\underline{24},0),
\end{equation}
where
\begin{eqnarray}
(\underline{1},0) &=& (1,1,0,0), \\ 
(\underline{10},1) &=& (3,2,1/6,1) + (3^*,1,-2/3,1) + (1,1,1,1), \\ 
(\underline{10}^*,-1) &=& (3^*,2,-1/6,-1) + (3,1,2/3,-1) + (1,1,-1,-1), \\ 
(\underline{24},0) &=& (1,1,0,0) + (8,1,0,0) + (1,3,0,0) + (3,2,-5/6,0) + 
(3^*,2,5/6,0).
\end{eqnarray}
As the $(1,1,0,0)$ component of the $(\underline{24},0)$ acquires a vacuum 
expectation value at the grand-unification scale, the 45 generators of 
$SO(10)$ are reduced to the 12+1 generators of the SM plus $U(1)_\chi$, 
with exactly 32 would-be Goldstone bosons provided by the $(\underline{10},1) 
+ (\underline{10}^*,-1)$ components of the \underline{45} and the 
$(3,2,-5/6,0) + (3^*,2,5/6,0)$ components of the $(\underline{24},0)$. 
This means that $U(1)_\chi$ may survive to near the electroweak symmetry 
breaking scale.  It is also orthogonal to $U(1)_Y$, unlike recent proposals 
where $U(1)_{B-L}$ is used \cite{k08,hkor08,ko08,adrs08,bbms08,bmt09,kkr09}.  
In fact, the $U(1)_\chi$ charge is given by 
\begin{equation}
Q_\chi = 5(B-L) + 4Y = 5(B-L) + 4T_{3L} - 4Q.
\end{equation}
The neutral fermion singlet $N^c$ in the \underline{16} of $SO(10)$, often 
referred to as the right-handed neutrino, has $B = 0$, $L = -1$, and $Y = 0$, 
so it has $Q_\chi = 5$.  Similarly, $(u,d), u^c, e^c$ have $Q_\chi = 1$ and 
$(\nu,e), d^c$ have $Q_\chi = -3$.

To allow for quark and lepton masses, a Higgs scalar doublet
\begin{equation}
\Phi = (\phi^+,\phi^0) \sim (1,2,1/2,-2)
\end{equation}
under $SU(3)_C \times SU(2)_L \times U(1)_Y \times U(1)_\chi$ is needed, 
with the Yukawa interactions
\begin{equation}
(u \phi^0 - d \phi^+)u^c, ~~~ (d \bar{\phi}^0 + u \phi^-)d^c, ~~~ 
(e \bar{\phi}^0 + \nu \phi^-)e^c, ~~~ (\nu \phi^0 - e \phi^+)N^c.
\end{equation}
To break $U(1)_\chi$ and not the SM gauge group, a Higgs scalar $\eta$ 
transforming under only $U(1)_\chi$ is needed.  If $\eta$ has 
$Q_\chi = \pm 10$, then $N^c$ gets a Majorana mass and the usual seesaw 
mechanism is operable.  However, another more interesting choice is 
available, as shown below.

\noindent \underline{\it Additional singlets}~:~ Instead of one scalar with 
$Q_\chi = \pm 10$, two scalars transforming under $U(1)_\chi$, namely
\begin{equation}
\eta_1 \sim 1, ~~~ \eta_2 \sim 2,
\end{equation}
will be used.  In addition, neutral fermion singlets
\begin{equation}
S_3 \sim -3, ~~~ S_2 \sim 2, ~~~ S_1 \sim -1,
\end{equation}
are added.  Note that the set of one $S_3$, four $S_2$, and five $S_1$ is 
anomaly-free, because $(-3) + 4(2) + 5(-1) = 0$ and 
$(-27) + 4(8) + 5(-1) = 0$.  As a result, the Yukawa couplings
\begin{equation}
N^c S_3 \eta_2^\dagger, ~~~ S_3 S_2 \eta_1, ~~~ S_2 S_1 \eta_1^\dagger, ~~~ 
S_1 S_1 \eta_2,
\end{equation}
are allowed, as well as the scalar interaction terms
\begin{equation}
\eta_1^2 \eta_2^\dagger, ~~~ (\eta_1^\dagger \eta_1)^2, ~~~ 
(\eta_2^\dagger \eta_2)^2, ~~~(\eta_1^\dagger \eta_1)(\eta_2^\dagger \eta_2).
\end{equation}
Altogether, it is clear that the choice of particle content allows a 
multiplicatively conserved lepton parity to be defined, so that 
$\nu, e, e^c, N^c, S_3, S_1, \eta_1$ are odd and $S_2, \eta_2$ are even.  
The $U(1)_\chi$ gauge symmetry is broken by $\langle \eta_2 \rangle \neq 0$, 
whereas $\langle \eta_1 \rangle = 0$.

In the basis spanned by $\nu, N^c, S_3, S_1$, the $4 \times 4$ mass matrix 
is then given at tree level by
\begin{equation}
{\cal M} = \pmatrix{0 & m_D & 0 & 0 \cr m_D & 0 & m_N & 0 \cr 0 & m_N & 0 & 0 
\cr 0 & 0 & 0 & M_1}.
\end{equation}
In the above, $S_1$ gets a Majorana mass at the $U(1)_\chi$ breaking scale, 
but it decouples from the other three fields.  The remaining $3 \times 3$ 
submatrix is exactly of the form of Eq.~(2) without $m_{R,L}$.  The next 
step is to show that the latter are not zero but small, because they will be 
generated radiatvely.

\noindent \underline{\it Dark matter}~:~ Before showing the specific radiative 
mechanisms responsible for $\epsilon_{R,L}$, i.e. $m_{R,L}$ renamed, an 
important bonus of this proposal is the occurrence of dark-matter candidates, 
i.e. $S_2$ and $\eta_1$.  They have odd $R$ parity, i.e. $R = (-)^{3B+5L+2j}$, 
whereas all other particles have even $R$ parity.  This is another example of 
the possibility of generalized lepton number \cite{klm09,m09-2}.

\noindent \underline{\it Radiative masses}~:~ At tree level, $S_3$ links with 
$N^c$ to form a Dirac fermion with mass $m_N$ and $S_1$ gets a Majorana 
mass $M_1$, both at the scale of $U(1)_\chi$ breaking due to $\langle \eta_2 
\rangle$.  This leaves $S_2$ massless, but it picks up a radiative Majorana 
mass in one loop, as shown in Fig.~1.
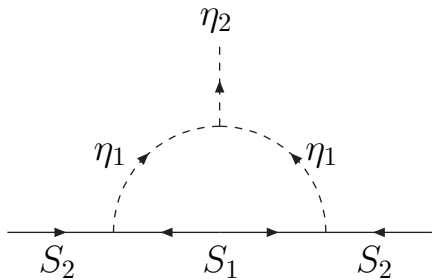
\begin{figure}[htb]
\begin{center}
\begin{picture}(500,100)(120,45)
\ArrowLine(270,50)(310,50)
\ArrowLine(350,50)(310,50)
\ArrowLine(350,50)(390,50)
\ArrowLine(430,50)(390,50)
\Text(290,33)[b]{\large$S_2$}
\Text(410,33)[b]{\large$S_2$}
\Text(352,33)[b]{\large$S_1$}
\Text(310,75)[b]{\large$\eta_1$}
\Text(390,75)[b]{\large$\eta_1$}
\Text(350,126)[b]{\large$\eta_2$}
\DashArrowLine(350,90)(350,120){3}
\DashArrowArcn(350,50)(40,180,90){3}
\DashArrowArc(350,50)(40,0,90){3}
\end{picture}
\end{center}
\caption[]{One-loop $S_2$ mass.}
\end{figure}
This is exactly analogous to the one-loop mechanism for neutrino mass first 
proposed in Ref.~\cite{m06}.  It is easily calculable from the exchange of 
Re($\eta_1$) and Im($\eta_1$) and is given by
\begin{equation}
m_2 = {f_{12}^2 M_1 \over 16 \pi^2} \left[ {m_R^2 \over m_R^2 - M_1^2} \ln 
{m_R^2 \over M_1^2} - {m_I^2 \over m_I^2 - M_1^2} \ln {m_I^2 \over M_1^2} 
\right].
\end{equation}
This means that $S_2$ is lighter than either Re($\eta_1$) or Im($\eta_1$), so 
the lightest $S_2$ should be a dark-matter candidate.  However, if $m_R < M_1$, 
Re($\eta_1$) is also stable, and similarly for Im($\eta_1$).  This is thus a 
natural scenario for the coexistence of several particles together as 
dark-matter candidates \cite{cmwy07}.

Once $S_2$ gets a mass, $S_3$ also gets a Majorana mass, as shown in Fig.~2.
\begin{figure}[htb]
\begin{center}
\begin{picture}(500,100)(120,45)
\ArrowLine(270,50)(310,50)
\ArrowLine(350,50)(390,50)
\ArrowLine(350,50)(310,50)
\ArrowLine(430,50)(390,50)
\Text(290,33)[b]{\large$S_3$}
\Text(410,33)[b]{\large$S_3$}
\Text(352,33)[b]{\large$S_2$}
\Text(310,75)[b]{\large$\eta_1$}
\Text(390,75)[b]{\large$\eta_1$}
\Text(350,126)[b]{\large$\eta_2$}
\Text(350,50)[]{\large$\times$}
\DashArrowLine(350,120)(350,90){3}
\DashArrowArc(350,50)(40,90,180){3}
\DashArrowArcn(350,50)(40,90,0){3}
\end{picture}
\end{center}
\caption[]{Two-loop scotogenic $S_3$ mass.}
\end{figure}
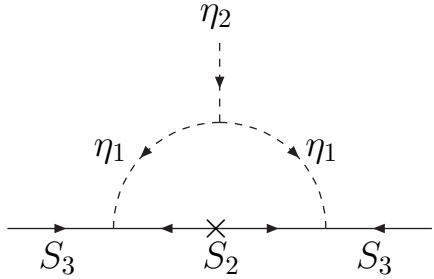
This is the $\epsilon_L$ term being sought after, and since it is a two-loop 
effect ($m_2$ itself being a one-loop effect), it is guaranteed to be small, 
as promised.  It is also a scotogenic mass, i.e. induced by darkness, 
because $S_2$ and $\eta_1$ have odd $R$, as pointed out previously.

There is also a mass term linking $S_3$ with $S_1$, as shown in Fig.~3.
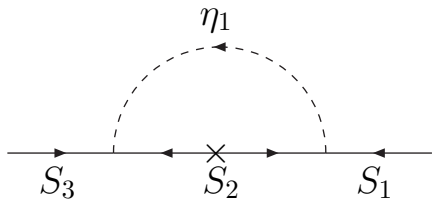
\begin{figure}[htb]
\begin{center}
\begin{picture}(500,80)(120,45)
\ArrowLine(270,50)(310,50)
\ArrowLine(350,50)(310,50)
\ArrowLine(350,50)(390,50)
\ArrowLine(430,50)(390,50)
\Text(290,33)[b]{\large$S_3$}
\Text(410,33)[b]{\large$S_1$}
\Text(352,33)[b]{\large$S_2$}
\Text(350,96)[b]{\large$\eta_1$}
\Text(350,50)[]{\large$\times$}
\DashArrowArc(350,50)(40,0,180){3}
\end{picture}
\end{center}
\caption[]{Two-loop mass linking $S_3$ with $S_1$.}
\end{figure}
Together with $M_1$, this gives a seesaw contribution to $\epsilon_L$ as well, 
but its magnitude is clearly much smaller than the $S_3 S_3$ term.
As for $\epsilon_R$, i.e. the $N^c N^c$ term, it is a three-loop effect 
and safely negligible. 

\noindent \underline{\it $U(1)_\chi$ phenomenology}~:~ With 
$\langle \phi^0 \rangle = v$ and $\langle \eta_2 \rangle = u$, both the $Z$ 
of the SM and $Z'_\chi$ become massive, but there is also $Z-Z'_\chi$ mixing 
which is of order $v/u \sim M_Z/M_{Z'_\chi}$.  Precision electroweak 
measurements at the $Z$ resonance constrain this mixing to be very small.  
To satisfy it without making $u$ very large, a second Higgs scalar doublet 
may be added, i.e. $\Phi' \sim (1,2,1/2,2)$ with $v'=v$, in which case 
$Z-Z'_\chi$ mixing is zero, and $M_{Z'_\chi}$ is not constrained except by 
the direct production of $Z'_\chi$.  The present experimental limit 
\cite{cdf07} is 822 GeV at 95\% CL.

Note that because of its $U(1)_\chi$ charge, $\Phi'$ does not couple to 
quarks or leptons, thus avoiding the appearance of flavor-changing neutral 
currents. It also does not link $\nu$ with $S_3$ or $S_1$ in Eq.~(20). 
Note further that the quartic $\eta_2^\dagger \eta_2^\dagger \Phi^\dagger \Phi'$ 
term is allowed, so that the introduction of $\Phi'$ does not create an extra 
global U(1) symmetry, thus avoiding the appearance of an unwanted massless 
Goldstone boson in the presence of $v'$.

If $Z'_\chi$ is not much heavier than 1 TeV, it will be discovered at the 
Large Hadron Collider (LHC), due to start taking data soon in 2009.  The 
key to verifying the radiative inverse seesaw mechanism is that $N^c$ must 
combine with $S_3$ to form a pseudo-Dirac fermion $N$ with lepton number $L=1$, 
as shown in Eqs.~(5) and (6).  If $M_{Z'_\chi} > 2 m_N$, then $Z'_\chi$ will 
decay into $N \bar{N}$ with subsequent decays $N \to e^- W^+, \nu Z$ and 
$\bar{N} \to e^+ W^-, \bar{\nu} Z$, etc.  This differs from the usual 
$U(1)_\chi$ expectation for $Z'_\chi \to N^c \bar{N}^c$ because $N^c$ is 
Majorana in that case.  Hence there would be both $e^\mp e^\mp W^\pm W^\pm$ and 
$e^\pm e^\mp W^\pm W^\mp$ final states.  The absence of the former would 
be the first indication of the inverse seesaw.  In addition, the 
branching-fraction ratio $B(Z'_\chi \to N \bar{N})/B(Z'_\chi \to e^+ e^-)$ 
is 17/5, whereas $B(Z'_\chi \to N^c \bar{N}^c)/B(Z'_\chi \to e^+ e^-)$ is 5/2.

The smoking gun of the scotogenic origin of $\epsilon_L$, i.e. the $S_3 S_3$ 
term, is the decay $N \to \bar{S}_2 \eta_1^\dagger$, which is invisible. 
This would be very difficult to ascertain at the LHC, but in a future 
possible linear $e^+e^-$ collider, $Z'_\chi$ may be produced at resonance. 
In that case, $Z'_\chi \to N \bar{N}$ with $N \to \bar{S}_2 \eta_1^\dagger$ 
and $\bar{N} \to e^+ W^-$ or $\bar{\nu} Z$ would provide the proof 
necessary.

\noindent \underline{\it Conclusion}~:~  The origin of neutrino mass may 
well be the inverse seesaw mechanism, i.e. $m_\nu \simeq m_D^2 \epsilon_L 
/m_N^2$.  To understand the possibility of $m_N \sim 1$ TeV and $\epsilon_L 
\sim 10$ keV, an extra $U(1)_\chi$ gauge symmetry is proposed, from the simple 
breaking of $SO(10) \to SU(3)_C \times SU(2)_L \times U(1)_Y \times U(1)_\chi$ 
by a single Higgs multiplet transforming as \underline{45} of $SO(10)$. 
With the addition of fermion and scalar singlets, $m_N$ comes from the 
breaking of $U(1)_\chi$.  As a bonus, dark-matter candidates emerge which 
are responsible for generating $\epsilon_L$ in two loops.  This is the 
first example of a radiative inverse seesaw mechanism, which is verifiable 
at the TeV scale.  It allows for observable unitarity violation of the 
$3 \times 3$ neutrino mixing matrix, as well as lepton flavor violation.

\noindent \underline{\it Acknowledgment}~:~ This work was supported in part 
by the U.~S.~Department of Energy under Grant No.~DE-FG03-94ER40837.

\bibliographystyle{unsrt}

\end{document}